\documentclass{article}
\usepackage[inline]{enumitem}
\usepackage{epsfig}
\usepackage{listings}
\usepackage[hyphens]{url}
\usepackage{hyperref}
\usepackage{subcaption,todonotes}
\usepackage{rotating}
\usepackage{booktabs,multirow,amssymb}
\usepackage[inline]{enumitem}

\usepackage{color}
\usepackage{authblk}
\definecolor{pblue}{rgb}{0.13,0.13,1}
\definecolor{pgreen}{rgb}{0,0.5,0}
\definecolor{pred}{rgb}{0.9,0,0}
\definecolor{pgrey}{rgb}{0.46,0.45,0.48}

\definecolor{gray}{rgb}{0.4,0.4,0.4}
\definecolor{darkblue}{rgb}{0.0,0.0,0.6}
\definecolor{cyan}{rgb}{0.0,0.6,0.6}

\date{}
\begin{document}
\title{Knock-Knock: The unbearable lightness of Android Notifications}
\author{Constantinos Patsakis and Efthimios Alepis}
\affil{Department of Informatics, University of Piraeus,\\ 80, Karaoli \& Dimitriou str.,18534, Piraeus, Greece\\
\protect{\url{kpatsak@unipi.gr}}, \protect{\url{talepis@unipi.gr}}}
\maketitle

\abstract{Android Notifications can be considered as essential parts in Human-Smartphone interaction and inextricable modules of modern mobile applications that can facilitate User Interaction and improve User Experience. This paper presents how this well-crafted and thoroughly documented mechanism, provided by the OS can be exploited by an adversary. More precisely, we present attacks that result either in forging smartphone application notifications to lure the user in disclosing sensitive information, or manipulate Android Notifications to launch a Denial of Service attack to the users' device, locally and remotely, rendering them unusable. This paper concludes by proposing generic countermeasures for the discussed security threats.}

\section{Introduction}
Modern mobile devices have penetrated our daily lives in the past years, making them an indispensable part of our daily lives. Initially, their goal was to provide communication to users via various means e.g. calls, short messages, video conference and chatting. Nevertheless, currently, they have stormed other activities such as infotainment, Internet browsing, fitness monitoring, or even finance. The fact that these devices are small enough to be seamlessly carried on daily basis and the fact that they are able to perform multiple tasks and process data from a plethora of embedded sensors has enabled developers to create numerous applications and a new niche market.

The interaction between users and mobile devices increases over the years and as a result the means of realizing this bi-directional communication has also evolved. Back in 2014 a comprehensive study on mobile phone notifications \cite{Pielot:2014:ISM:2628363.2628364} revealed that users had to deal with 63.5 notifications on average per day, mostly from messengers and email. Obviously,  this number has been increased significantly over the last years with the rise of apps such as WhatsApp and Facebook messenger and notification messages overpassing billions in daily basis \cite{whatsapp}.

At the same time recent studies reveal that push notifications draw users' attention, with user average opening notification rates being over 90\% \cite{biznessapps}. Notifications also allow developers to increase user engagement with their app and improve user retention rates \cite{biznessapps}. Mobile notifications seem to clearly influence user engagement positively and also improve user conversion rates. As stated in \cite{localytics}:
\begin{quote}
``{\it In 2015, users who enabled push notifications launched an app an average of 14.7 times per month, whereas users who did not only launched an app 5.4 times per month. In other words, users who opted in to push messages averaged 3x more app launches than those who opted out.}''\end{quote}
 Equivalent findings are also reported in \cite{urbanairship}: \begin{quote}``Analysis of 63 million app users' first 90-days reveals more frequent messaging increases mobile app retention rates by 3X to 10X''.
\end{quote}

In view of the above this paper illustrates the necessity of providing mobile users with secured and trusted mechanisms for their interaction with mobile devices. Towards this end, this paper analyzes the basic interaction mechanisms available in Android, the worlds' most popular mobile platform to date. Our research illustrates flaws in the Android's notification mechanisms which can lead to a number of attacks, presented in this work. Subsequently, this paper also discusses about countermeasures in order to provide defense solutions to the attacks.

\noindent{\bf Main contributions:} The main contribution of this work is to illustrate successful attacks in the most recent versions of Android, using AOSP as a reference. These attacks can affect additively the majority of Android users to date. More specifically, this paper presents forging notification attacks that include home-screen shortcuts, attacks in notifications that result in DOS and also attacks concerning web push notifications. To the best of our knowledge, and according to the malware samples of \cite{wei2017deep}, notifications are used for aggressive advertisement via the malware families of Airpush and Kuguo. Overall, we aim in providing a thorough analysis in the core means of realizing Human-Smartphone interaction, both for providing a way to detect flaws and also for better locating their corresponding working solutions.

\noindent{\bf Organization of this work:}
The rest of this work is organized as follows. In the next section we present the related work. Section \ref{sec:hsi} provides the background and the basis for the problem setting. Then, Section \ref{sec:notifattacked} presents two distinct categories of attacks on Android Notifications. Finally, the article concludes discussing possible countermeasures.

\section{Related work}
\label{sec:related}
Android was designed to run in devices with constrained capabilities in terms of both computation or size. The size of the device, as Android mainly targets handheld devices, implies many restrictions in the UI. As a result, the UI can be considered as a set of layers which are stacked one on top of the other. However, as highlighted in \cite{chin2011analyzing}: ``{\it Phishing attacks can be mounted convincingly because the Android UI does not identify the currently running application.}''. This was exploited by \cite{niemietz2012ui} who managed to create the first UI redressing attack for Android. While drawing on top of other activities and partially covering them, researchers have recently started to find novel ways to do it. The bulk of the attacks exploit a recently introduced Android permission, the \texttt{System\_Alert\_Window} \cite{Ying:2016:ADA:2897845.2897897,fratantonio17:cloakdagger}. According to Google Developer resources \cite{system_alert}: ``{\it Very few apps should use this permission; these windows are intended for system-level interaction with the user}''. Nevertheless, due to backward compatibility issues, an adversary may easily use this permission without the user's knowledge or consent by targeting lower API levels in the app. While the attacks have rather severe impact, the apps that can exploit this feature are rather limited and can be easily found by simply scanning the manifest for the corresponding permissions. On the contrary, a more stealth attack allows an adversary to overlay activities without requesting any permission from the user \cite{alepisRAID2017}.

However, in order to timely present the user a screen which requires him to provide his credentials, one needs to be aware of which is the foreground activity. Methods to do this are discussed in \cite{chen2014peeking,alepisRAID2017}, however, they are rendered useless as of Nougat, since access to \texttt{/proc} has been significantly shrinked.
To counter this lack of knowledge, an adversary may resort to other means, e.g. masquerade as a legitimate app and convince the user to interact. Note that all apps are aware of which apps are already installed in the device, so the adversary can easily find a target one. In this regard, in \cite{xu2012abusing}; a closely related work to us, a set of attacks which exploit notifications was proposed. The concept is that the user has been tricked into installing a malicious app named Notish which issues notifications that look like ones from other legitimate apps luring them to disclose sensitive information e.g. credentials. The attacks that the authors demonstrate apply not only to Android, but iOS and Blackberry, while a spam scenario also exists.

Perhaps the most relevant to ours can be considered the work of \cite{xu2012abusing}, for this reason it is further analyzed. Back in 2012 the authors where supporting that they first presented a paper regarding the security of notifications in mobile phones. Specifically regarding Android, they covered platforms 2.3 and 4.0 (API level 14). Nevertheless, many years have passed and considering the small mobile OS lifetime, this period was quite significant for many reasons. Most importantly, Android has changed a lot since 2012, providing 12 newer API levels to date, with the introduction of Android Oreo, always hardening its security. To this end, notification services can be considered as much different compared with the situation in 2012. More specifically, not only are their security mechanisms more improved, but also the attacks described in \cite{xu2012abusing} do not actually apply in the settings of Android in its latest versions. The most basic reason is that these attacks depended on hard-coded resource graphics, which, after being submitted for publication to Google Play Store, they would immediately be discarded by the Google Bouncer as it would find them fraudulent. Moreover, most of the notifications' functionalities that are exploited in this paper, such as replacing the notifications' icon, where introduced very recently. e.g. the ``Notification.Builder setSmallIcon (Icon icon)'' function was added in API level 23. Furthermore, in the last versions of Android, the app name has been supplementarily added in notifications, enhancing their security even more, once again rendering past attacks useless.

The interested reader may refer to \cite{felt2011phishing,virvilis2014mobile} for more on phishing attacks on Android.

\section{Human-Smartphone Interaction}
\label{sec:hsi}
Analyzing the interaction between humans and smartphones, we may come up in itemizing the most profound and basic reasons for humans using this ``special'' computing device, namely the smartphone, that has drown much of the users' interest over the last decade. The categorization of these actions could inarguably differ, facilitating other points of view, however, for the purposes of this study we split them in the following four categories:
\begin{enumerate*}[label=(\alph*)]
	\item Communicating with others (calling, texting etc.)
	\item Internet browsing,
	\item Using 3rd party applications (m-banking, infotainment etc.),
	\item Responding to notifications
\end{enumerate*}
To conclude to these four basic categories of actions in using a mobile device, we  investigated why, how, when and under what conditions an average user might be using a smartphone. While there might be overlaps in these categories, discussed in the following paragraphs, we consider these reasons of realizing human-smartphone interaction as distinct.

Communicating with others via calling, texting etc. are basic and fundamental actions performed in mobile devices even before the existence of smartphones, namely since the appearance of feature phones. This kind of interaction is accomplished through applications that accompany the OS and are developed by the OS vendor. Certainly, both phone calling and text messaging can be also realized by third party applications, while VoIP solutions are also rapidly appearing. Nevertheless, this can be considered as a fundamental reason for using a mobile device, since its very existence and before having these devices being able to connect to the Internet.

Using the well-known mobile browsers for visiting and interacting with web pages through mobile devices is also considered as a very important reason of human-smartphone interaction. Notably, in 2016 mobile phone users who visit the WWW overpassed the corresponding number of personal computer users \cite{statcounter}. The latter not only highlights the importance of such an interaction, but it also indicates the closer connection to humans' lives that smartphones have managed to acquire.

Using third party applications in a smartphone can be considered as one of the most important and basic reasons of interaction between users and mobiles. The incorporation of all kinds of applications that are of the users' interests has been perhaps the basic reason for smartphones having operating systems that support this kind of functionality and thus make them considerably distinguishable to ``ordinary'' feature phones. The extraordinary adoption of app stores where users can browse and install applications of numerous categories supports the aforementioned argument. Additionally, recent studies \cite{flurry} between smartphones provide strong indications that mobile applications are the users' preferred way of interaction when using their smartphone for some reason (e.g. play a game, buy tickets, check a personal bank account balance), compared with the corresponding services relying in web pages. The usage of third party applications of course implies the existence of notifications in a high percentage of use cases, such as receiving and consequently responding to instant messages. This part of interaction, however, is covered separately in the following paragraphs.

Finally, as already mentioned, we consider that responding to notifications is a special interaction between humans and smartphones. By using the term ``responding'', we consider both the cases when users actually respond to received notifications by ``opening'' them, as well as  the cases of even more basic actions, such as using the smartphones' built-in notification drawer to see and/or read the messages of the notifications, even if the users, in some cases, decide not to ``open'' a notification and just ``clear'' it after reading part of its message. This kind of interaction can be considered of great importance as it forms a very common reason of using the smartphone in modern human-smartphone interaction and because it is expected to grow even more and become more important in the years to come. A smartphone can receive notifications for a large number of reasons that include, yet are not limited to, the following:
\begin{itemize}
	\item Having missed a phone call
	\item A newly received text message
	\item System automatically updated applications
	\item System automatically updated applications
	\item Results from periodically scheduled jobs such as software update checking and virus scanning
	\item Messages from carriers
	\item Push notifications coming from external resources, corresponding to installed applications. Perhaps the most notable ones in this category are instant messaging applications such as WhatsApp, Viber and Facebook messenger
	\item Web push notifications that constitute a quite ``recent'' addition in most mobile browsers' capabilities, where even not used web pages are able to transmit their notifications to targeted mobile devices
	\item Local app notifications, where third party apps create and issue a notification in order to be read by the user, or require an action taken by the user
\end{itemize}

The aforementioned categories of notifications are significant and basic, however one should investigate the underlying reasons why this ``kind'' of interaction is important more thoroughly. Thus the reader may have both a rational explanation and also the evidence deriving from the users' experience about the authors' claims. Notifying in terms of mobile computing means, at its fundamental definition, implies finding a way to reach the user, gain his/her attention. This can be variously achieved, aligned with the Operating System's supported corresponding functionalities. A foreground application being used by a smartphone user can change/update its Graphical User Interface (GUI) to provide new information to the user. The same use-case may occur in a web environment, where a web page may dynamically adjust its content. When considering applications that are not currently active, or are running in the background, or are even closed since the mobile device is switched off, the ways to accomplish user notification changes significantly, since additional parameters have to be taken into account. Notably, new information that would change the contents of an Android activity but would not be launched until a user actually opened the corresponding application cannot be considered as a timely, nor an acceptable way to realize user attention. On the contrary, modern mobile notifications try to ``force'' user interaction and do not rely on waiting when or whether the user decides to check their corresponding app.

In this sense, native mobile applications may choose among a variety of programmatically feasible solutions in order to draw users' attention, such as newly launched activities, opened web pages through browsers, dialog messages, toast messages and of course native Android notifications. The latter, however, has some considerable advantages to count, as it will be further explained, thus may be chosen as the prevailing solution in a majority of use cases where an application is not being used, or when a device is switched off.

Essentially, all the aforementioned ``solutions'' can work towards the direction of informing the user about something. Nevertheless, as it will be further discussed, in the cases where users are either not using their mobile device, or a specific app, developers may opt in favor of the native Android notifications to accomplish user-app interaction. Toast messages involve two basic drawbacks in these cases. First, they are useless when a mobile device is switched off or locked, since they will not appear. In the case where the mobile device is unlocked and awaken, a displayed toast message can only provide short information to a user, for only 5 seconds without necessarily providing the identity of the issuer. Newly launched activities, or launched mobile pages through mobile browsers are definitely more ``permanent'' than toast messages solutions, since they do not disappear after a specific period of time, nor get affected by the state of the mobile device, namely when they are closed and/or locked. However, both are invasive in terms of user-mobile interaction since they impose their presence as the foreground app in the mobile device's main User Interface (UI). Moreover, there is no guarantee that they are going to be the foreground app when the user unlocks the phone, since other, newer, activities might have been launched, putting them in the background. Android dialog messages suffer the same disadvantages too. Additionally, dialog messages are required to hold the quite ``dangerous'' system permission that allows them to draw over other screens to accomplish the desirable result, namely the SYSTEM\_ALERT\_WINDOW permission, that can be maliciously used \cite{fratantonio17:cloakdagger}.

Deductively, we lead to reason why the Android notifications are the preferred, and the suggested by Google, way of realizing the communication between a user and an unused app, or even more precisely, between a user and an application that is not in the foreground. In addition, the internal design of Android notifications provides them with some valuable assets in terms of establishing an effective and accepted solution for asynchronous or semi-synchronous background initiated communication between users and apps. These assets include great levels of effectiveness in terms of OSes resources usage, permanence and user friendliness in terms of providing a noninvasive way of notifying users. Nevertheless, there are two points that require significant attention. All kinds of notifications that are issued from a ``background'' process when either a device is locked or when a user is using an irrelevant application also means that the identity of the notification issuer is also very critical. When a user is actually using an application and the application's content changes, users can feel quite sure, presumably, that the changed content originates from the application they use. On the contrary, when users receive incoming information from an application they are not actively using, they rationally need to be able to verify its actual source. This is the reason why, when receiving an email from an unknown source requesting the bank credentials to proceed to an action issued by a well-known bank, most users, hopefully, consider this email as fake and subsequently delete it. Accordingly, a mobile phone user is expecting to be able to confirm a notifications' issuer true identity before proceeding to an action that could range from posting an unwanted message to a social network, to exposing the user's login credentials for his/her bank account, or a company's server login. To this end, the notifications' nature can be considered as even more deceptive when they are asynchronous to the users' active interaction, since by definition are not expected to appear when users are using the issuing applications and respectively know their origin.

As already mentioned, the aim of this paper is twofold. One the one hand, provides evidence, regrettably, that Android notifications can be provably be insecure in contrast to their wide adoption and increased interest by both developers and users. On the other hand, the authors also suggest solutions that may address the arising security issues. Towards this direction, even though applying the countermeasures will not provide ground proof that Android notifications will subsequently become secure, closing security holes still improves them and also helps towards the direction of maturing an ever developing and constantly evolving mobile operating system. For these reasons, the analyzed underlying security issues and not only statically illustrated, rather than the causes of their origin are investigated and generic solutions as countermeasures are proposed. Leaving the programming level and anatomizing the more abstract level of the Android Notifications' infrastructure in terms of Human-Smartphone interaction not only reveals this ``mechanism's'' profound architecture, but also projects both its strengths and weaknesses.

\section{Attacks on Android Notifications}
\label{sec:notifattacked}
In this section we discuss attacks on Android Notifications. More specifically, our main focus will be to illustrate the feasibility of forging application notifications, which expose the  users' privacy and security. Secondly, we present how Android's notifications can be exploited to launch a Denial of Service (DoS) attack, both locally and remotely.

\subsection{Forging app notifications}
\begin{figure*}
	\centering
	\includegraphics[width=.7\textwidth]{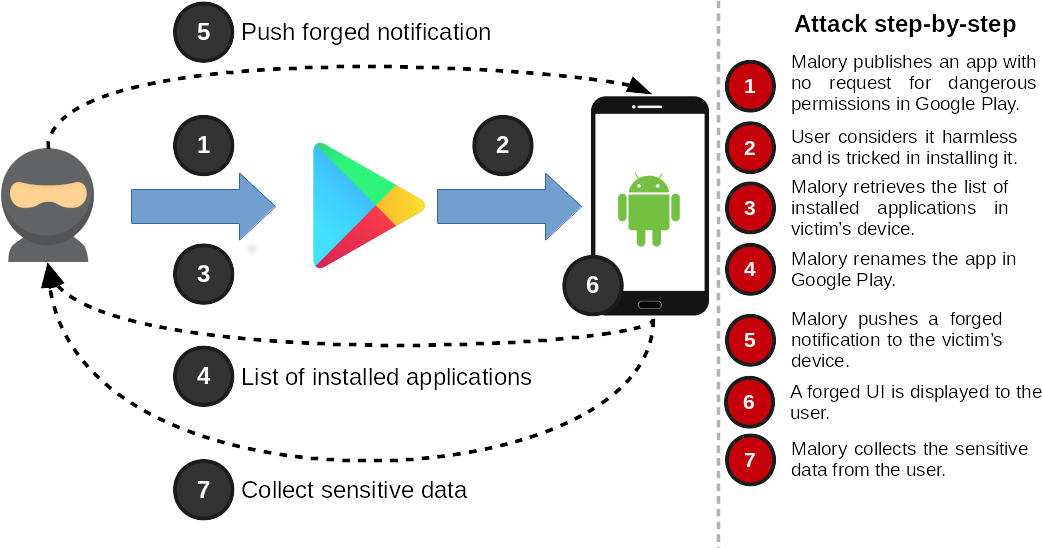}
	\caption{Attack overview.}
	\label{fig:attack_overview}
\end{figure*}

As already analyzed in the previous section, the interaction between smartphone apps and users is bidirectional. In the case of users' initiated interaction, namely user-to-app, users are having knowledge about the applications that they launch and use. In this paper we are going to prove that by designing a fine-tuned, yet based of evidence, attack to Android users, the app-to-user interaction can be exploited. In particular, the attack that is illustrated in this section involves a number of steps from the attacker's side, involving fundamental Android components and services. The impact of forging notifications can be considered rather high, as it can be used to deceive them to collect user credentials, perform user profiling, or even blackmailing. The steps of the attack are illustrated in Figure \ref{fig:attack_overview}. Following, a use case is also presented and analyzed.
For the purposes of highlighting the dangers that accompany this attack, in our step-by-step use case scenario we have chosen ``PayPal'' as the ``target'' app. The steps are the following:
\begin{itemize}
\item A user installs a zero-permission app through Google Play. The apps name is BobApp and requires only Internet access (even this permission can by bypassed if necessary).
\item The installed app retrieves the list of installed applications in the victim's device. The list is communicated to a service owned by the attacker.
\item The attacker determines whether an app that he would like to make an attack to is available in victims' devices. In this scenario it is ``PayPal''.
\item The attacker issues an update for the app, through Google Play, where the app's name is replaced, namely ``BobApp'' becomes ``PayPal''. The update is expected to be launched automatically, usually by night, when the device is unattended (e.g. probably left charging and connected to a Wifi).
\item After a successful application update, the malicious app's name has been changed, while the user (owner) has no way to know about it.
\item To build a complete notification, the malicious app requires a title, a text and also the target app's icon. This is accomplished by utilizing the actual genuine target app's resources. More specifically, the target app's package is located and the app's graphics are retrieved through the application's resources and the application's metadata. As a result, a new notification is triggered, with an identical to the genuine app interface.
\item After triggering the malicious notification, the user is expected to select it and subsequently launch a malicious activity. The malicious activity can further utilize the genuine app's resources in order to provide a UI that will lure the user.
\item Finally, the user is asked for some private info (credentials) compromising her/his account.
\end{itemize}

\begin{figure}
	\includegraphics[width=.45\columnwidth]{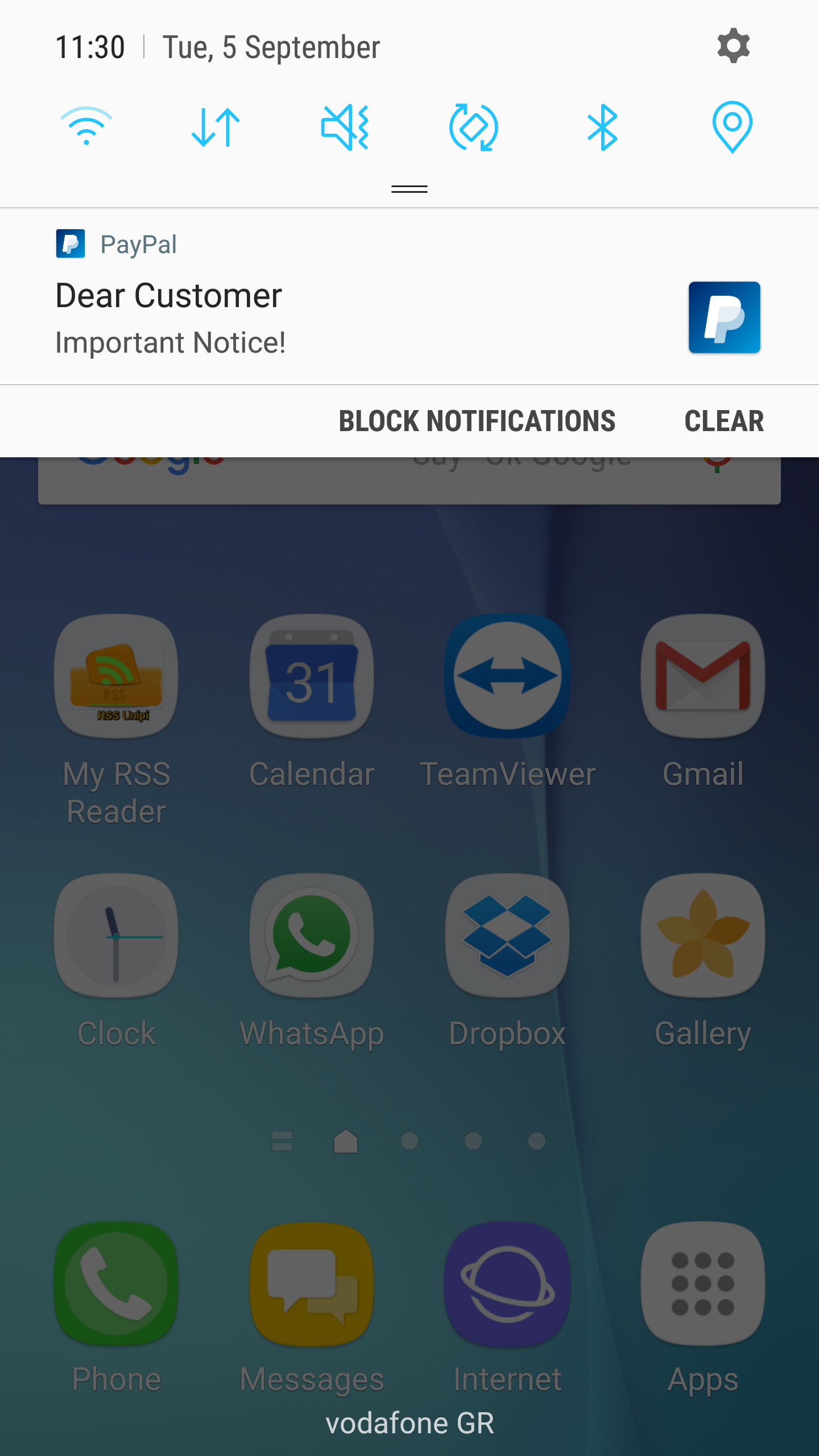}
	\includegraphics[width=.45\columnwidth]{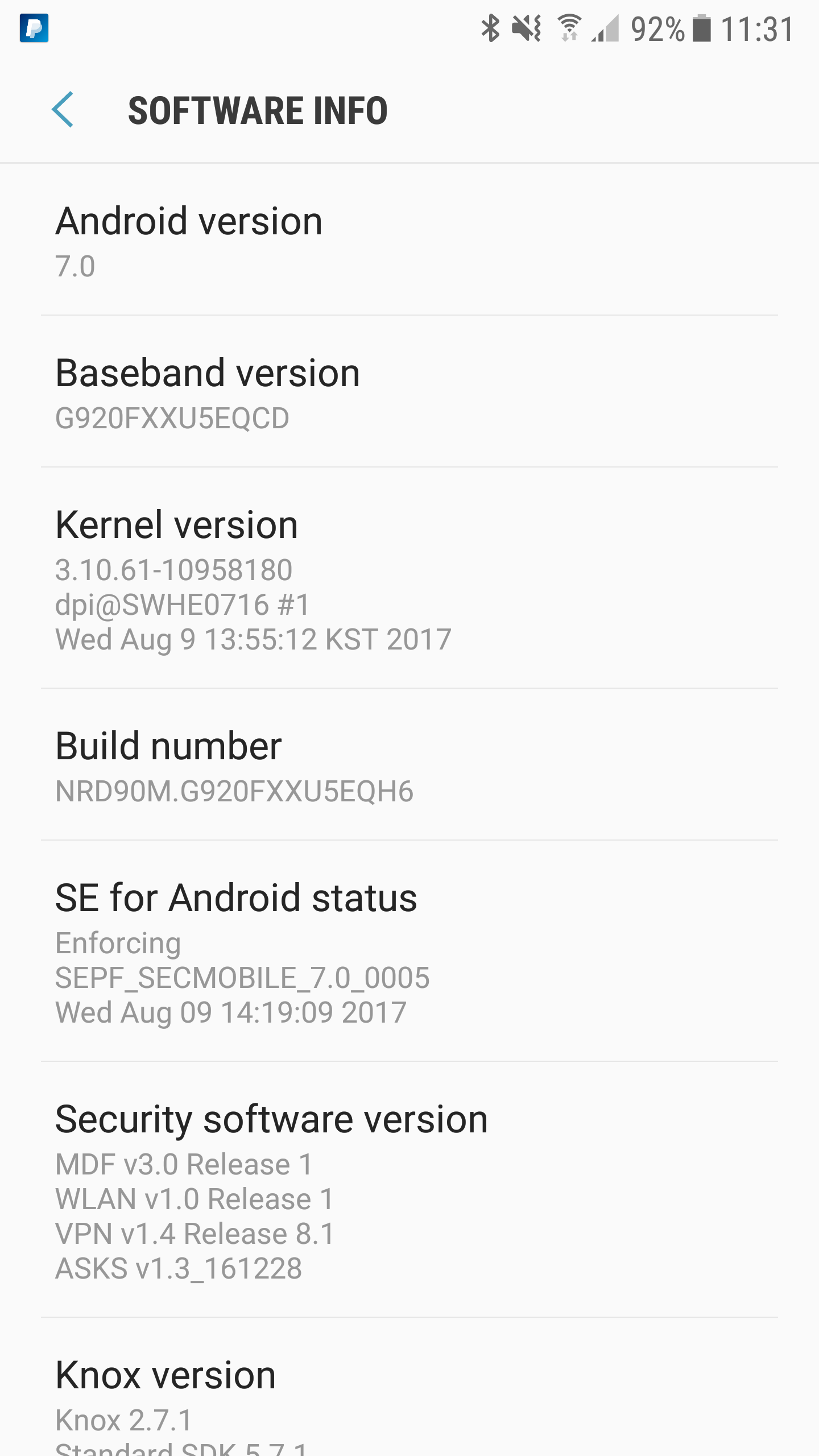}
	\caption{A forged notification from another app.}
	\label{fig:forged_not}
\end{figure}

The above scenario has been tested both locally and through Google Play. A forged notification for PayPal on Nougat is illustrated in Figure \ref{fig:forged_not}.  More interestingly, since the ID for an app located in Google Play consists only of its package name, one can very easily find numerous apps in Google Play sharing the same name. Moreover, after thorough research we have come up with the conclusion that bypassing a notification's actual app name is almost impossible by other means in AOSP. Achieving this programmatically requires the ``substitute\_notification\_app\_name'\'' signature permission, which only 2-3 system apps actively have. Indeed, our research revealed that these apps are ``Easter Eggs'', ``Google Play'' and ``Shell''. The combination of using other apps' resources and changing the app name arbitrarily through background silent updates, makes the described attack scenario both effective and real. Indeed, the described attack proves to the readers that Android users can be led to a situation where they would not be able to reason about the origin of their smartphones' notifications.

Finally, another part of the Android's user interaction mechanism that has been found to have flaws is the home-screen application shortcuts. Home-screen shortcuts are coupled to notifications in many cases since once one or more notifications have been issued by an application, an indicative change appears in the corresponding app's shortcut (e.g. an indicating number of unread notifications). Nonetheless, home-screen shortcuts are also being frequently used by users to initiate an interaction with an app. Our independent research has revealed that Android home-screen app shortcuts can be easily forged and a malicious application can appear on a device's home-screen as another application, with identical icon and name. Both the home-screen icon and also the name of the shortcut are not hard-coded and do not originate from the app's resources. As a result every application is able to create a home-screen shortcut as being another app. In this sense, forged app notifications can ``cooperate'' with forged app home-screen shortcuts to further deceive the user. Even without having notifications issued, home-screen shortcuts provide attackers with another attack-vector in the human-smartphone interaction.

Both security issues, regarding forged notifications and forged home-screen shortcuts, have been responsibly disclosed to Google's Android Security Team.

\subsection{DoS through Notifications}
Following another approach, notifications can be used to launch a denial of service both locally and remotely. The attack exploits a bug in NotificationManager when allocating memory for creating a Notification. The notification's builder object expects a specific size of the icon, yet allocates memory for any given image graphic. Potentially, this bug may allow arbitrary code execution, however, to this point, we were not able to execute it. As a result, the attack; currently being patched by Android Security Team, is launched when a properly crafted notification is sent to the NotificationManger class. After launching the attack, the System UI process repeatedly crashes blocking the user from making any other interaction apart from answering a call and rebooting. Notably, while answering calls is allowed, the UI does not revert to the original state.  Furthermore, by registering a broadcast receiver Android object waiting for the ``BOOT\_COMPLETED'' event and re-issuing the malicious notification makes the device unusable, since the device will immediately fall to the previous looping state. Possible countermeasures, before a patch is applied by Google, is to uninstall the malicious application through a possible ``safe mode'' device state, or re-flashing the whole device.

To launch this attack, actually almost no code is required, since only a ``big'' image of high resolution is needed to be sent to NotificationManager for rendering. In our tests we used a high resolution  (4096x4096) PNG file, of rather low file size (2.79 KB), however larger graphics have the same effect. Such a ``malicious'' image can either be stored in the application's resources, or be loaded dynamically. Nevertheless, some critical issues arise by its actual capability to be able to be loaded dynamically, from other resources than the actual application that is firing the notification.

Clearly, the above scenario is triggered locally, by a malicious installed app. Nonetheless, this can also be triggered remotely. As of API level 23, the Notification.Builder class includes a new method, namely ``Notification.Builder setSmallIcon (Icon icon)'', which accepts an icon rather that a resource. This way, many apps (e.g. Youtube), use Internet resources to download their graphics. Since this information is not private and  servers want to take advantage of caching, such graphics are mostly transmitted via plain HTTP. Using a simple man-in-the-middle attack, an adversary can replace the requested graphic with a high resolution graphic and brick the devices remotely.

Even when examining the case where a device is targeted in API levels less than 23, where the new ``setSmallIcon'' function was not available and even if there was the possibility of pre-checking the resources of each app that could be used as icons, the problem still exists. Namely, since apps' resources are ``public'' to other installed apps, a malicious application could very easily scan the device for all installed apps' resources and select a high resolution image and use it for the attack.

\subsection{Web push Notifications}
Recently, the ability to fire notifications has been also given to web pages, with a big number of modern and popular mobile browsers supporting this feature. In our tests we have successfully tested web push notifications on Android devices running Chrome version $\geq$ 42, Mozilla version $\geq$ 44 and Samsung Internet browser version $\geq$ 4.0. Having the ability to notify a user asynchronously was one of the very basic advantages that native mobile applications had in the past, in contrast to web pages, whose lifetime of interaction with the user was bounded to the time the user spent in browsing on a specific web page.

Web push notifications were introduced to fill this gap. This functionality is established through a ``bridge'' between the native app world and the web pages ecosystem, provided by the mobile browser. A mobile browser is a ``special'' kind of software entity. Since it operates in both of these ``worlds'', it is actually a native mobile application installed in a smartphone, while simultaneously its basic purpose is to serve web pages. As a result, having some special permissions given by the user explicitly, a web page is able to send a push notification asynchronously to the browser and consequently, the browser is responsible for ``communicating'' this notification to the user's device, utilizing the OSes native mechanisms.

However, the latter introduces another attack vector for phishing attacks. Assume the case of a user accepting notifications from a malicious web page from his mobile browser. The web page then can push an arbitrary notification to the device at any given point. Practically, in our phishing scenario, the malicious web page pushes a notification with the icon of an app with millions of downloads, expecting users to respond. The notification redirects the user to a webpage which replicates the UI of the targeted app requesting for sensitive information, e.g. credentials. While the notification may verbally state that the notification originates from the browser, yet the visual identity, the displayed icon and the notification text, may lure many users.
Under the precondition that a web page is able to determine which applications a user has installed in her/his device, recently published in \cite{eye2017}, a forging app notification use case could be able to appear in web push notifications too. Both text and graphics can be easily arbitrarily be loaded through web resources. However, the actual notifications ``issuer'' in these cases is always the name of the browser (e.g. Chrome), accompanied by the notification's title. In this sense, a user may be lured and open a malicious notification believing that one of her/his installed mobile apps has fired it by seeing the graphic and the text involved. However, as already mentioned, a more attentive examination by the user could reveal that the notification was fired by the browser and not by the actual installed app. Nonetheless, since a web notification containing information about an installed app is not very common, users may more easily be deceived in such a scenario.

The case of issuing a DoS attach to a device through web push notifications has been also investigated. However, in our experiments, this kind of attack was unsuccessful. The graphic is probably firstly ``rendered'' by the browser to meet the proper size, thus blocking the attack. Processing a very high definition graphic by a mobile app would require processing time, which could eventually lead to an Application not Responding (ANR) situation. The browsers in our tests seem to properly handle these situations and result in either firing the notification with its graphic scaled, or providing a ``default'', ``harmless'' icon for the notification to be issued.

\section{Discussion and Conclusions}
\label{sec:countermeasures}
Due to size constraints, Android UI lacks the source verification of graphic components, exposing the users to many risks. In the literature, several approaches have been proposed to counter such issues. For instance, a third party framework named ``SecureView'' was proposed in \cite{xu2012abusing}. SecureView allows the user to choose a security image as well as writing a text-based security greeting after installing an application in her/his device. This way, whenever a sensitive view is displayed, the application can show the security image and greeting on the sensitive view to provide view authentication to the user. These kinds of countermeasures can also be found to have drawbacks. One the one hand, using such a framework from applications implicitly means trusting a third party company. On the other, having users supply both different security images and greetings for a number of their installed applications might not work for obvious reasons, including user frustration and negative user experience.

In \cite{bianchi2015app} the researchers propose the introduction of a visual identity to facilitate the user identify which app he is actually interacting. Their solution tracks the origin of the app that created the displayed dialog and presents it in the notification bar.
Wu et al. monitor the WindowManagerService to determine the presence of a floating window by re-calculating the Z-order of all windows and hooking all the calls which are triggered when creating and clicking on a window \cite{wu2016analysis}. This approach may not counter UI replication attacks, but it defends many overlay attacks. Ying et al. propose a similar visual identity to Bianchi et al. for identifying the source of a UI element also tweaking Window Manager \cite{ying2016attacks}. WindowGuard hooks the Activity Manager, the Window Manager and Package Manager services to monitor overlaying UI elements and activity transition in order to detect possible attacks \cite{ren2017windowguard}.

An automated screenshot mechanism is proposed in both \cite{malisa2016mobile,malisa2017detecting} to find similarities between apps and determine whether a UI attack is being made by an app.

Other solutions to UI attacks are presented in \cite{fernandes2016android,marforio2016evaluation,wu2016effective}.
For more on phishing attacks and countermeasures the interested reader may refer to \cite{heartfield2016taxonomy,aleroud2017phishing}

From the aforementioned defense mechanisms it is clear that only some of the proposed mechanisms can provide only partial measures against our attacks. The main reason is the context awareness of our app and the renaming of the app. In this regard, the displayed UI is rendered according to the targeted applications that are already installed in the device, so screenshot mechanisms are rendered useless. A very important aspect that needs to be taken into consideration is that the paper's described attacks are actually triggered by the user. Consequently, there are no overlays to be detected by the system and due to the renaming, all the appropriate visual signatures can be easily circumvented.

Therefore, when the interaction between human and smartphone is initiated by the user, then it is of utmost importance to ensure that the user is actually launching the application s/he intended to launch. While using the applications' basic shortcuts it seems that s/he is protected, since an application using another apps graphic as a the app's launcher icon requires it being placed in the apps resources, which consequently would give Google Bouncer the ability to ``intercede'' when another apps graphic has been detected in Google Play. Unfortunately, this is not the case with home-screen app shortcuts. As already discussed, home-screen app shortcuts can be easily arbitrarily created by other apps, producing ``identical'' forged app shortcuts. In this case, the OS must provide some new rules and/or checks to overpass this problem e.g. test whether a specific icon matches the resources of another app.

In the cases where the interaction between human and smartphone is implicitly initiated by applications, the problem seems a little more complicated. As it is already mentioned, some ``ways'' that evolve in user notification, such as activity launches, dialogs and toasts, have clear disadvantages and are rejected by developers in most cases. The major contribution of this work is the investigation of the Android Notifications' mechanism from its security perspective. Generically, notifications are not tightly coupled to their issuer, since they refer to a way of ``leaving a notice'' for the user through the OSes features, both enhancing user experience and also preserving the valuable smartphone resources. Consequently, this notice is left to be opened by the user end navigate her/him to another UI to continue her/his interaction, presumably a native mobile app, or even a web page. It may be considered as ``common sense'' that anyone who ``picks up'' a notice in either her/his mailbox or her/his mobile navigation drawer, should be able to identify the sender. As a result, all involved parties in software development should work towards this direction, safeguarding the users, as their ultimate aim.

Having these in mind, regarding the notifications' issues, there are several solutions that could be proposed for the OS vendor. Being able to prove the notification's issuer id, would involve an id to be passed either to the notifications current app name, or the the required graphic or even to both. As we have proven both the name of the notification and also the graphics could be easily being forged. The apps' package name can be considered as a candidate that identifies each app, which also cannot exist as duplicate in Google Play. Nevertheless, it has some drawbacks, since while enhancing users' security, it negatively affects user experience. Forcing the icons/graphics of notifications originating only from local resources is another option, which would nevertheless require the OS to rollback to a previous solution, with clear negative results in the market. Other kinds of side-countermeasures could include removing the potential from apps to being able to determine which apps are installed in users' devices, or making a special check for Google Play apps who are making a change in their app name and consequently removing them for the automated updated app list.

The presented work has resulted in four responsibly disclosed to Google security issues that are not public yet at the time of writing. More specifically, two issues regard the described methods in forging app notifications in all Android versions from Marshmallow to Oreo, one issue regarding forging home-screen shortcuts, affecting all Android versions and one issue regarding the illustrated DoS attack, affecting all Android versions. The authors have followed all the guidelines regarding the disclosure of the issues, both in terms of following the security procedures and also in terms of giving the company enough time to be able to prepare patches.

\section*{Acknowledgments}
\label{sec:Acknowledgments}
This work was supported by the European Commission under the Horizon 2020 Programme (H2020), as part of the \emph{OPERANDO} project (Grant Agreement no. 653704) and is based upon work from COST Action \emph{CRYPTACUS}, supported by COST (European Cooperation in Science and Technology). The authors would like to thank \emph{ElevenPaths} for their valuable feedback and granting them access to Tacyt.

\bibliographystyle{plain}
\bibliography{refs}

\end{document}